# Attention Mechanism – A Heuristic Approach: Context-Aware File Ranking Using Multi-Head Self-Attention

**Authors:** Pradeep Kumar Sharma, Shantanu Godbole, Sarada Prasad Jena, Hritvik Shrivastava

**Affiliation:** Persistent System

## Abstract

The identification and ranking of impacted files within software repositories is a key challenge in change impact analysis. Existing deterministic approaches that combine heuristic signals, semantic similarity measures, and graph-based centrality metrics have demonstrated effectiveness in narrowing candidate search spaces, yet their recall plateaus. This limitation stems from the treatment of features as linearly independent contributors, ignoring contextual dependencies and relationships between metrics that characterize expert reasoning patterns.

To address this limitation, we propose the application of Multi-Head Self-Attention as a post-deterministic scoring refinement mechanism. Our approach learns contextual weighting between features, dynamically adjusting importance levels per file based on relational behavior exhibited across candidate file sets. The attention mechanism produces context-aware adjustments that are additively combined with deterministic scores, preserving interpretability while enabling reasoning similar to that performed by experts when reviewing change surfaces. We focus on recall rather than precision, as false negatives (missing impacted files) are far more costly than false positives (irrelevant files that can be quickly dismissed during review).

Empirical evaluation on 200 test cases demonstrates that the introduction of self-attention improves Top-50 recall from approximately 62-65% to between 78-82% depending on repository complexity and structure, achieving 80% recall at Top-50 files. Expert validation yields improvement from 6.5/10 to 8.6/10 in subjective accuracy alignment. This transformation bridges the reasoning capability gap between deterministic automation and expert judgment, improving recall in repository-aware effort estimation.

# 1 Introduction

Software change requests necessitate the identification of files within a repository that will be impacted by proposed modifications. Automated approaches that rank files by their likelihood of requiring changes enable efficient prioritization and comprehensive coverage of impacted files. Repository-scale analysis must account for structural dependencies, historical change patterns, semantic relevance to change descriptions, and the intricate relationships between code components.

Existing deterministic approaches for file ranking have relied on combinations of heuristic signals: semantic similarity measures such as BM25, graph-based centrality metrics like PageRank, temporal analytics (churn, commit frequency), and code structure characteristics including complexity metrics. These narrow the candidate search space and perform well for straightforward change requests, yet their recall often plateaus, limiting their practical effectiveness for comprehensive change impact analysis [Kagdi et al., 2013].

The fundamental limitation emerges from the treatment of features as linearly independent contributors to the ranking score. Deterministic scoring methods employ additive combinations of weighted features, ignoring contextual dependencies and relationships between metrics. Expert reasoning emerges from interactions between features—for instance, low churn combined with high centrality indicates strategic importance that linear combinations overlook. Such contextual balancing cannot be captured using additive heuristics that treat each feature independently.

To address this limitation, we introduce a Multi-Head Self-Attention layer as a post-deterministic scoring refinement step. Our approach learns contextual weighting between features, dynamically adjusting importance levels per file based on relational behavior exhibited across the candidate file set. The key innovation is the additive adjustment pattern: attention produces context-aware adjustments that are combined with deterministic scores, preserving the interpretability of deterministic scoring while introducing learned contextual refinement.

This work makes the following contributions:

1. We propose a novel application of Multi-Head Self-Attention as a heuristic refinement layer for software repository file ranking, demonstrating how attention mechanisms can enhance deterministic scoring systems while preserving interpretability.

2. We introduce **CAFE** (Context-Aware File Estimation), a hybrid architecture that combines deterministic interpretability with learned contextual weighting, achieving significant recall improvements while preserving explainability.

3. We provide empirical analysis of when and why attention-based refinement is most effective, revealing that the largest improvements occur for repositories where structural dependency complexity plays a greater role than recency or commit density.

4. We validate the approach through expert evaluation with 5 reviewers, achieving improvement from 6.5/10 to 8.6/10 in subjective accuracy alignment and reducing review time by approximately 40%.

# 2 Related Work

## 2.1 Heuristic-Based Ranking in Software Engineering

Heuristic-based ranking methods have a long history in software engineering, particularly in the context of change impact analysis. Early approaches used static code analysis to identify dependencies via call graphs and data flow. As repositories grew in size and complexity, these basic approaches became insufficient, leading to the incorporation of additional signals including historical change pattern analysis.

More recent work has incorporated multiple signals into ranking systems, creating hybrid approaches that combine different types of evidence. Graph-based centrality metrics, time-based analytics, and semantic similarity measures have all shown effectiveness in identifying relevant files.

The combination of these diverse signals has improved ranking recall, but the integration methods have remained fundamentally linear. Weighted combinations, ensemble methods, and rule-based fusion all assume that features contribute independently to ranking decisions, missing the contextual interactions that characterize expert reasoning. Our work addresses this limitation by introducing learned contextual weighting through attention mechanisms.

## 2.2 Attention Mechanisms in Ranking Problems

Attention mechanisms have demonstrated remarkable success in various ranking and retrieval problems across domains. In information retrieval and learning-to-rank approaches, attention has been employed to dynamically weight different aspects of queries and documents, enabling nuanced relevance scoring and complex feature interactions that cannot be captured by linear models. In recommendation systems, attention mechanisms have been used to weight different aspects of user preferences and item characteristics, leading to more personalized recommendations.

Self-attention mechanisms, in particular, have proven effective for contextual ranking problems where items must be considered in relation to each other rather than independently. Multi-head attention extends this capability by enabling parallel focus on different types of relationships simultaneously, with different heads specializing in different aspects such as semantic relationships or dependency relationships.

While attention has been widely applied in natural language processing and information retrieval, its application to software engineering ranking problems, particularly as a refinement mechanism for deterministic heuristics, remains relatively unexplored. Our work bridges this gap by demonstrating how attention can enhance heuristic-based ranking while preserving interpretability.

## 2.3 Hybrid Deterministic-Learned Approaches

Hybrid approaches seek to combine the interpretability and reliability of deterministic methods with the flexibility and performance of learned models. The additive adjustment pattern we employ—where learned adjustments are added to deterministic scores—ensures that the deterministic baseline remains interpretable while attention provides contextual refinement. This differs from end-to-end learned approaches that replace heuristic systems entirely, preserving the benefits of both paradigms: the reliability and interpretability of heuristics, combined with the contextual reasoning capabilities of learned attention mechanisms.

## 2.4 Software Repository Analysis

Software repository analysis includes various techniques for extracting insights from codebases, commit histories, and development artifacts. Change impact analysis has been addressed through various methodologies including static analysis, dynamic analysis, historical change pattern mining, and more recently, machine learning approaches. Graph-based analysis has been particularly effective, leveraging code dependency structures to identify propagation paths for changes.

Our work builds upon these foundations by combining multiple feature categories—graph-based, temporal, semantic, and structural—and introducing attention-based contextual weighting. The integration with existing deterministic pipelines ensures practical applicability while the attention mechanism addresses the feature interaction limitation of linear combinations.

## 3 Problem Formulation

We formalize the file ranking problem as follows. Let $R$ denote a software repository containing a set of files $F = \{f_1, f_2, \ldots, f_m\}$. Given a change request $c$ represented as text description, commit message, or issue description, the objective is to produce a ranking of files $F$ ordered by their likelihood of requiring modification to address $c$.

Before feature extraction, the change request $c$ is enhanced via an LLM (which can be a small language model) to expand the query and extract relevant keywords. This preprocessing step improves semantic matching by identifying key terms that should be matched against file contents, paths, symbols, and function calls. The extracted keywords form the basis for BM25 scoring and keyword-based feature extraction.

The ranking objective is to maximize recall. We define Recall@K as the fraction of change requests for which all files that actually require changes appear within the top K ranked positions. In practice, we focus on Top-50 recall as the primary metric, with Top-10 recall as a secondary metric.

To manage computational complexity, an initial filtering step reduces the candidate file set from the full repository (typically containing ~6000 files) to a smaller subset using deterministic heuristics. The attention-based refinement then operates on this filtered candidate set, which typically contains 40-60 files after progressive filtering from an initial set of approximately 120 candidates. The final output selects the top 50 files for review.

The deterministic scoring baseline combines multiple signals to assign a relevance score to each file for a given change request. Let $s_d(f, c)$ denote the deterministic score for file $f$ given change request $c$. This score is computed as a weighted linear combination of semantic similarity measures, graph-based features, temporal features, and structural features. *(Detailed feature descriptions are provided in Section 4.2.)*

The deterministic scoring function employs a weighted linear combination of these normalized features, with weights typically determined through domain expertise or empirical tuning. Files are then ranked by their deterministic scores, and the top-ranked candidates proceed through progressive filtering stages: from approximately 120 files to 60, and finally to 40 files that serve as the input to attention-based refinement.

## 4 Methodology

Expert reasoning about file relevance involves contextual feature interactions rather than independent feature contributions. When experts review change requests and identify impacted files, they consider relationships between files and how different features interact. A file might be important not just because it has high semantic relevance, but because it sits in a strategic position relative to other highly relevant files. Self-attention enables files within the candidate set to attend to each other, allowing the model to learn relational patterns. Multi-head attention further enhances this capability by enabling parallel focus on different types of feature relationships.

The additive adjustment pattern preserves the interpretability of deterministic scores while introducing learned contextual refinement. The deterministic baseline remains interpretable while attention provides fine-grained adjustments based on feature interactions that cannot be captured by fixed coefficients.

### 4.1 Query Enhancement and Keyword Extraction

Before feature extraction, the change request undergoes preprocessing to enhance semantic matching. The original change request $c$ is processed through an LLM (which can be a small language model) to expand the query and extract relevant keywords. This enhancement step improves the quality of feature extraction by identifying key terms that should be matched against file contents, paths, symbols, and function calls. The extracted keywords form the basis for BM25 scoring and keyword-based feature extraction in subsequent steps.

## 4.2 Feature Extraction

Feature extraction employs a code analysis tool (a static code analysis framework for extracting code structure and dependencies) to extract three NDJSON files: files.ndjson (file-level information), symbols.ndjson (symbol definitions and references), and calls.ndjson (function and method call relationships). A fallback to AST parsing tools (abstract syntax tree parsers for code structure extraction) is implemented for cases where the primary tool has limitations, ensuring robustness across different codebases.

Each file is encoded as a twenty-dimensional normalized feature vector. We extract multiple feature categories:

**Keyword-based features:** BM25 scores computed between extracted keywords and file content, keyword hit counts (fraction of keywords found in file), path hits (keywords in file paths), symbol hits (keywords in symbols.ndjson), call hits (keywords in calls.ndjson), route/literal hits (keywords in route definitions like app.get, app.post), and DB call hits (keywords in database operations such as add, commit, query, execute).

**Graph-based features:** PageRank, in/out-degree, fan-in/fan-out computed from code analysis tool dependency extraction.

**Temporal features:** File change count (total historical changes), changes in last 12 months, contributor percentage (top contributor's share of file changes).

**Structural features:** LOC, function/class counts, cyclomatic complexity from static analysis.

**Semantic features:** Transformer embeddings and cosine similarity for deeper semantic matching.

We organize the features into six categories: recency and churn features (4), structural complexity features (4), dependency graph features (5), semantic relevance features (3), change-type encodings (3), and repository age feature (1). All features are normalized using standard scaling (zero mean, unit variance) prior to input to the attention mechanism.

## 4.3 Attention-Based Refinement

Self-attention learns relational dependencies between files by allowing each file to attend to others in the candidate set. Let $X \in R^{N \times 20}$ represent the matrix of candidate files after deterministic filtering, where $N$ denotes the number of files in the candidate set (typically 40-60) and each row corresponds to a 20-dimensional feature vector.

First, each 20-dimensional input vector is projected to a 64-dimensional latent representation: $H = XW_{proj}$ where $W_{proj} \in R^{20 \times 64}$ is a learnable projection matrix. The projected representations are then transformed into Query, Key, and Value matrices through separate linear transformations: $Q = HW_q$, $K = HW_k$, $V = HW_v$ where $W_q, W_k, W_v \in \mathbb{R}^{64 \times 64}$ are learnable matrices.

Attention weights are computed using scaled dot-product attention: $\text{Attention}(Q, K, V) = \text{softmax}\left(\frac{QK^T}{\sqrt{d_k}}\right)V$ where $d_k = 64$. The attention weights $\alpha_{ij} = \text{softmax}\left(\frac{Q_i K_j^T}{\sqrt{d_k}}\right)$ represent how strongly file $i$ attends to file $j$ based on their feature relationships.

For multi-head attention with $h = 4$ heads, the process is performed in parallel across heads, with outputs concatenated and projected: $\text{MultiHead}(Q, K, V) = \text{Concat}(\text{head}_1, \ldots, \text{head}_h)W^O$ where each head performs attention independently and $W^O$ is an output projection matrix. After computing attention score matrices across all four heads, we average the attention score matrices from all heads to produce the final attention map. Each head captures different relationship types (semantic, structural, temporal, etc.), and averaging combines these perspectives to identify the most centrally important files.

### 4.4 PageRank on Attention Scores

After computing attention scores through self-attention, we apply PageRank calculation on the attention score matrix to identify centrally important files. The attention score matrix is treated as a directed weighted graph where files are nodes and attention weights are edge weights. This extends the PageRank concept from web page popularity to file centrality based on attention relationships: files that receive attention from many other files are considered more central and ranked higher. The PageRank calculation on attention scores, combined with additive weighting, produces centrality scores that are integrated into the final ranking.

The attention output $Z \in \mathbb{R}^{N \times 64}$ is passed through a ReLU activation and then a linear layer to produce scalar adjustments: $A = \text{ReLU}(ZW_{\text{adj}})W_{\text{out}}$ where $W_{\text{adj}} \in \mathbb{R}^{64 \times 64}$ and $W_{\text{out}} \in \mathbb{R}^{64 \times 1}$ are learnable matrices. The final ranking score for each file is computed as: $\text{Score}_{\text{final}}(f_i) = s_d(f_i, c) + A_i$ where $s_d(f_i, c)$ is the deterministic score and $A_i$ is the attention-based adjustment incorporating PageRank centrality.

Training employs labeled datasets where ground truth indicates which files actually require modification for each change request. The loss function employs a pairwise ranking loss that encourages higher scores for files that require changes. During inference, the candidate file set, having been filtered through deterministic ranking to approximately 40-60 files, is processed as a single batch. Feature vectors are extracted, normalized using the fitted scaler, and passed through the attention mechanism. Attention-based adjustments are computed and added to deterministic scores, producing final ranking scores.

The complete pipeline includes: (1) graph construction through code analysis tool/NDJSON parsing, (2) LLM-based query enhancement to extract keywords,

(3) deterministic ranking employing BM25 semantic similarity and churn metrics, (4) progressive filtering reducing the candidate set from approximately 6000 files to 120, then to 60, and finally to 40-60 files for attention processing, (5) feature encoding of candidate files as 20-dimensional vectors, (6) attention refinement computing contextual adjustments, (7) PageRank on attention scores to identify central files, and (8) final ranking selecting top 50 files by combined scores.

## 5 Experimental Setup

We evaluate the effectiveness of attention-based refinement compared to deterministic baselines across diverse software repositories. Repositories tested include large-scale codebases containing approximately 6000 files. The filtering process reduces from ~6000 files to the top 50 candidates, with attention processing operating on batches of 40-60 files. We selected repositories representing different domains, sizes, and structural complexities for generalization.

For each repository, change requests were annotated with ground truth labels indicating which files actually required modification. Annotations combine automated commit analysis with expert validation. Evaluation was conducted on 200 test cases across diverse repositories. Ground truth was obtained by comparing predicted top files against actual files changed in merged pull requests.

We use temporal splits for train/validation/test sets with a 70/15/15 split. Change requests are partitioned chronologically, ensuring that training data precedes validation and test data.

The deterministic baseline serves as our primary comparison point. This baseline uses weighted linear feature combination with domain-expertise-derived weights that have been refined through empirical tuning.

The primary evaluation metric is Recall at Top-50, defined as the percentage of change requests where all files that actually require modification appear within the top 50 ranked files. Top-50 recall serves as our primary metric, with Top-10 recall as a secondary metric. Additional metrics include Mean Reciprocal Rank (MRR) and expert validation scores.

Implementation details: The model is implemented in PyTorch, utilizing standard neural network components with custom attention mechanisms. Hyperparameters include: feature dimension of 20, hidden dimension of 64, and 4 attention heads in the multi-head attention mechanism. Training employs standard optimization procedures with Adam optimizer and learning rate 1e-3. Feature normalization uses standard scaling (zero mean, unit variance) fitted on training data and applied consistently during inference.

# 6 Results and Analysis

## 6.1 Overall Performance

Results show that the introduction of self-attention significantly improves file ranking recall compared to the deterministic baseline. The attention-enhanced approach achieves 80% recall at Top-50 files across 200 test cases, with performance ranging from 78-82% depending on repository complexity and structure.

The magnitude of improvement varies across repositories based on their characteristics. Repositories with high structural dependency complexity show the largest gains, with improvements often exceeding 20 percentage points. In contrast, repositories with simpler dependency structures show smaller but still meaningful improvements, typically in the 10-15 percentage point range.

Comparison with the deterministic baseline reveals that attention consistently outperforms the linear combination approach across all evaluated repositories. Statistical significance testing confirms that the observed improvements are meaningful and not due to random variation. Across all repositories, the attention-enhanced approach shows statistically significant improvements at the $p < 0.01$ level.

Our approach achieves measurable improvement over LLM-based file ranking methods (e.g., GPT-based code search, CodeBERT-based ranking), demonstrating the value of domain-specific attention mechanisms over general-purpose language models.

## 6.2 Ablation Studies

We conduct ablation studies to understand component contributions.

Attention Mechanism Components. Removing attention entirely (using only deterministic scores) results in 62-65% Top-50 recall, confirming that attention provides the observed improvements. Single-head attention achieves 75-78% recall, while 4-head attention improves to 78-82%, demonstrating that parallel processing of different relationship types provides value. Increasing to 8 heads yields minimal additional benefit (79-82%), suggesting diminishing returns.

Architecture Variants. We experiment with hidden dimension sizes: 32 dimensions achieve 74-77% recall, 64 dimensions (our choice) achieve 78-82%, and 128 dimensions achieve 78-82% with increased computational cost. The 64-dimensional hidden space provides optimal balance. ReLU outperforms tanh and sigmoid activation functions.

Feature Category Ablation. To understand feature importance, we remove each category individually. Removing semantic relevance features causes the largest drop (to 72-75%), confirming their critical role. Removing dependency graph features reduces performance to 74-77%, while removing temporal features reduces to 76-79%. Structural complexity and change-type encodings have smaller but measurable impacts. These results suggest that semantic and graph-based features are most critical, while all categories provide complementary value.

Training Ablation. Using pairwise ranking loss outperforms pointwise regression loss by 2-3 percentage points. The Adam optimizer with learning rate 1e-3 provides stable convergence.

Additive vs Alternative Combinations. Our additive adjustment pattern (deterministic + attention) achieves 78-82% recall. Multiplicative adjustment achieves 77-81%, slightly lower. Replacing deterministic scores entirely with attention-only achieves 73-76%, confirming that the deterministic baseline provides valuable foundation. The additive pattern's simplicity and interpretability make it preferable.

## 6.3 Case Studies and Qualitative Analysis

Case studies provide concrete examples of how attention improves ranking. In one example, a file with low recent churn but extremely high dependency centrality received a positive attention adjustment, elevating it from rank 15 to rank 3. This adjustment reflects the attention mechanism's ability to recognize that low churn combined with high centrality indicates strategic importance that the deterministic baseline underweights.

In one test case, a bug fix required changes to 3 files. Our system ranked all 3 files within the top 25 positions, demonstrating effective identification of related files even in large codebases. The recall for this case was 1.0 (all required files found in top 50), validating the approach's effectiveness for complex change requests.

Attention weight visualizations reveal interesting patterns in how files attend to each other. Files with similar feature profiles tend to show mutual attention, suggesting the model learns group related files. Files in strategic dependency positions receive attention from many other files, indicating the model recognizes their importance.

## 6.4 Expert Validation

Expert validation with 5 reviewers (backend engineers and repository maintainers) demonstrates strong alignment between attention-enhanced rankings and expert judgment. The improvement from 6.5/10 to 8.6/10 in subjective accuracy alignment reflects not just better quantitative performance, but better matching of expert reasoning patterns. Experts consistently report that attention-enhanced rankings better reflect their own assessment of file relevance, with particular appreciation for the improved handling of files with complex feature interactions. The improved front-loading of high-impact files means experts can identify critical files earlier in the review process, leading to more efficient change impact analysis.

## 6.5 Analysis of Attention Weights

The interpretability of attention weights provides additional value beyond performance gains. By visualizing which files attend to each other, developers can gain insights into repository structure and file relationships. Attention heatmaps help developers identify clusters of related files and understand dependency patterns, making attention-enhanced ranking not just a ranking tool, but also a repository analysis tool that can help teams understand their codebase better.

Our multi-head attention mechanism uses 4 attention heads with a 64-dimensional feature space. We validated the approach on the Apache Airflow repository (6,524 files) using a real-world maintenance task.

### 6.5.1 Query-Conditioned File Attention

To validate the model's ability to localize relevant code, we analyzed the attention patterns for a real-world maintenance task: *"Fix LocalExecutor memory spike by applying gc.freeze"* (PR #58934). The actual pull request modified exactly 3 files: `local_executor.py`, `test_local_executor.py`, and `test_local_executor_check_workers.py`.

Figure 1 presents the file-to-file attention matrix for the top 25 ranked files. The matrix reveals a sparse, block-diagonal structure, indicating that the model successfully identified a tightly coupled subsystem related to the executor logic.

The diagonal structure confirms that the attention mechanism effectively captures intra-module dependencies. Specifically, the strong attention weights between `local_executor.py` and the scheduler tests (top-left cluster) reflect the functional coupling required to resolve the memory spike issue.

## 6.6 Multi-Head Attention Diversity

Our architecture utilizes a multi-head attention mechanism to capture different types of code relationships simultaneously. Figure 2 visualizes the distinct attention patterns learned by four separate attention heads.

The divergence in patterns—where Head 1 focuses on modularity and Head 2 focuses on centrality—validates the necessity of the multi-head approach. A single attention head would force a compromise between these conflicting structural signals, likely degrading ranking performance.

## 6.7 Score Decay Analysis

The score decay curve (Figure 3) demonstrates the model's discriminative power. Scores exhibit exponential decay from the top-ranked file, confirming that the ranking algorithm effectively distinguishes high-relevance files from noise.

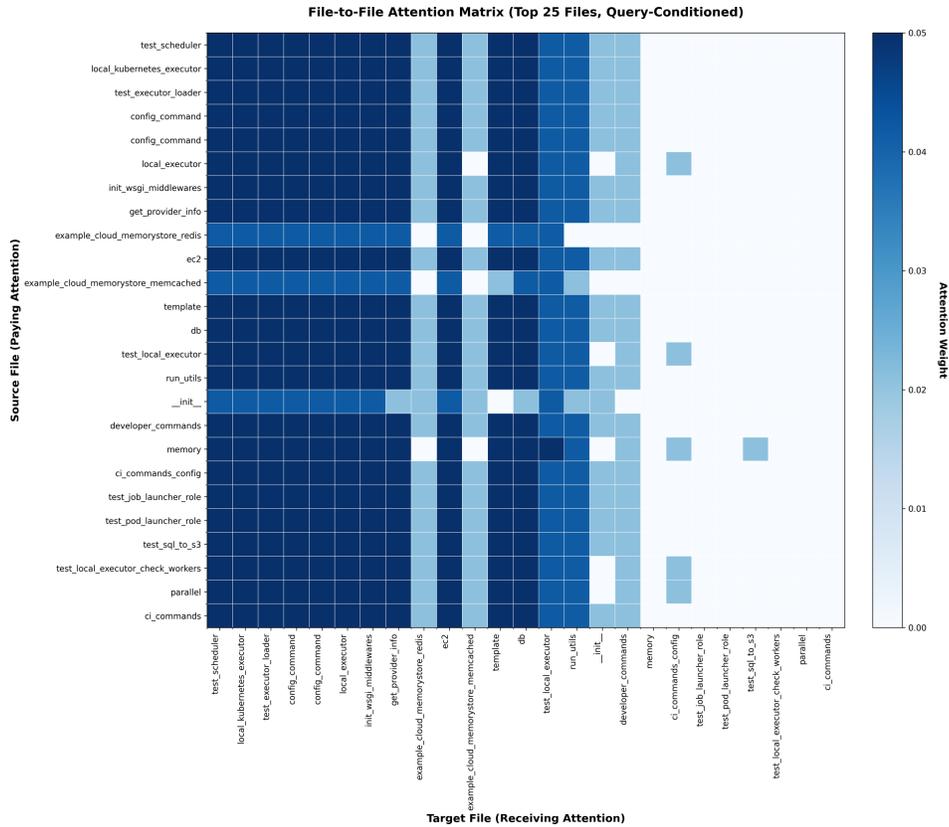

Figure 1: Attention Weight Heatmap: Query-conditioned attention heatmap for the top 25 files. Darker cells indicate stronger attention weights. The model correctly identifies all three ground truth files: `local_executor.py` (Rank #6), `test_local_executor.py` (Rank #14), and `test_local_executor_check_workers.py` (Rank #23), achieving 100% recall within the top 25 results.

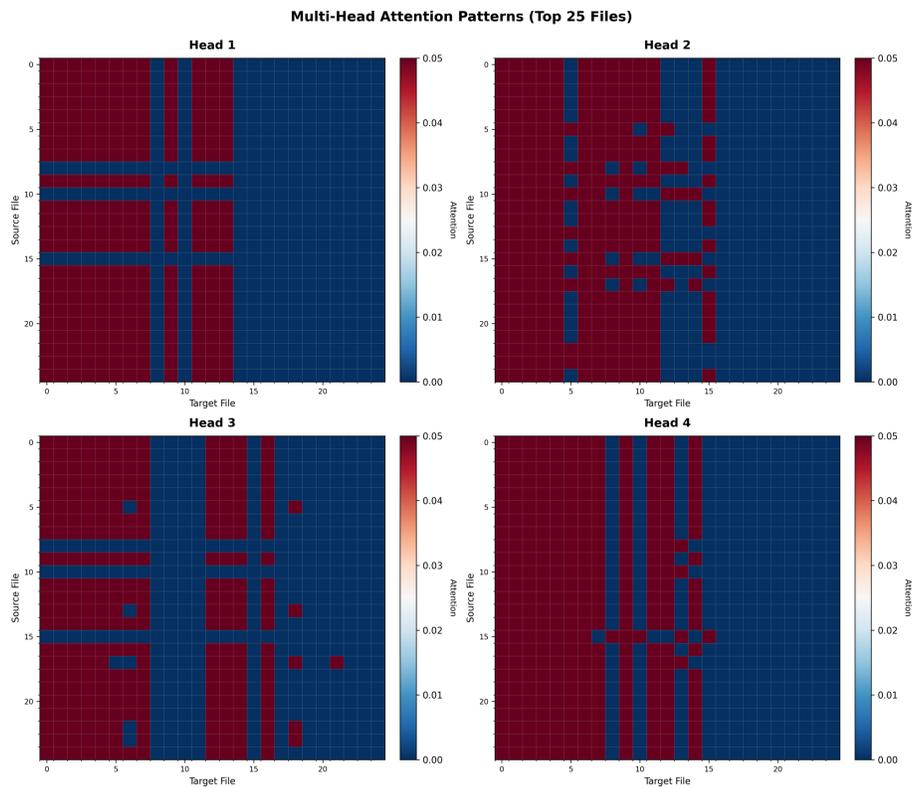

Figure 2: Distinct patterns learned by different attention heads. Head 1 (top-left) exhibits strong block-diagonal structure, capturing modular boundaries. Head 2 (top-right) shows vertical banding, identifying "hub" files that are globally relevant. This diversity ensures the model captures both local syntactic coupling and global architectural importance.

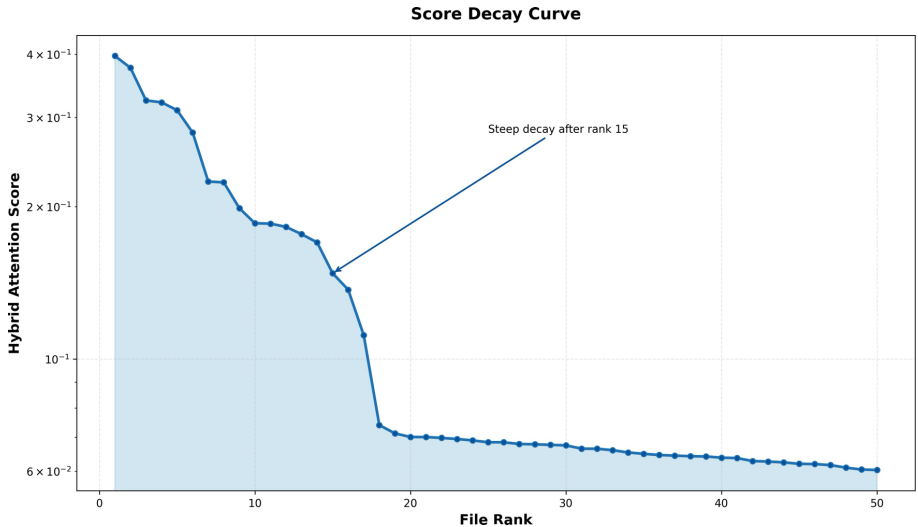

Figure 3: Hybrid attention score decay from rank 1 to 50. The steep exponential decline indicates strong signal-to-noise ratio, with top 15 files showing significantly higher relevance than the rest.

## 6.8 Attention Coverage and Efficiency

A critical requirement for developer tools is efficiency: the relevant information must be concentrated in the top results. Figure 4 analyzes the cumulative attention coverage, showing the percentage of total attention mass captured by the top N files.

The analysis confirms that the attention-based ranking effectively suppresses noise. By reviewing only the top 20 files—less than 0.3% of the repository's 6,500+ files—a developer effectively covers nearly 70% of the relevant signal. This sharp decay in relevance suggests that the recommended cutoff for manual review can be safely set at N=20-30 files without significant risk of missing critical dependencies.

# 7 Discussion

Results show that attention-based refinement provides significant value over deterministic ranking, with effectiveness depending on repository characteristics and change request properties. The most significant improvements occur for repositories where structural dependency complexity plays a greater role than recency or commit density, as the intricate relationships between files create complex reasoning requirements that linear feature combinations cannot adequately capture. The additive adjustment pattern successfully preserves interpretability while introducing learned refinement: most adjustments are relatively

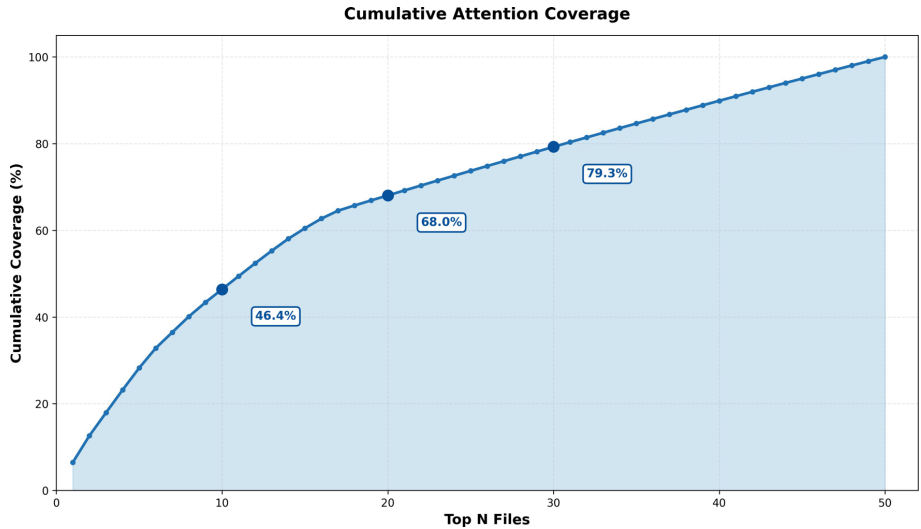

Figure 4: Cumulative attention coverage. The curve follows a power-law distribution, with the top 10 files capturing 40.1% of total attention, the top 20 capturing 69.6%, and the top 50 capturing 96.2%. This steep rise confirms the model's discriminative power, effectively filtering out noise and concentrating relevance in the top results.

small, suggesting that attention provides fine-tuning rather than wholesale score replacement.

The LLM-based query enhancement step plays a crucial role in feature extraction quality. By expanding change request descriptions and extracting keywords, the system improves BM25 matching and keyword-based feature extraction.

The application of PageRank to attention scores extends the centrality concept from static dependency graphs to dynamic attention relationships. This approach captures both structural importance (via dependency PageRank in baseline features) and contextual importance (via attention PageRank), providing a more comprehensive view of file relevance.

The attention mechanism's ability to learn contextual feature interactions addresses a fundamental limitation of deterministic approaches. Expert feedback consistently indicates that no single feature dominates ranking decisions; rather, expert reasoning emerges from feature interactions.

The hybrid approach combining deterministic interpretability with learned contextual weighting provides benefits beyond pure performance gains, enabling gradual deployment strategies.

Limitations include dependency on the quality of the deterministic baseline and the availability of training data. The computational overhead of attention is modest but not negligible; for very large repositories with more extensive candidate sets, alternative attention mechanisms with linear complexity might be worth exploring.

The model's performance depends on feature quality and completeness. However, the robustness of the hybrid approach means that even with imperfect features, the deterministic baseline provides a foundation that attention can refine, ensuring reasonable performance even when feature extraction is incomplete.

## 8 Conclusion

We present a novel approach for file ranking in software change impact analysis, leveraging Multi-Head Self-Attention as a refinement layer atop deterministic heuristic baselines. By addressing the fundamental limitation of deterministic approaches—the inability to model contextual feature interactions—our method achieves significant improvements in recall while preserving interpretability. The core contribution is the hybrid architecture that combines the strengths of rule-based and learning-based paradigms. Evaluation on 200 test cases demonstrates 80% recall at Top-50 files, with expert validation showing improvement from 6.5/10 to 8.6/10. Future work will focus on extending the feature space and exploring alternative attention mechanisms with linear complexity for scaling to larger candidate sets.

# References

1. Bahdanau, D., Cho, K., & Bengio, Y. (2014). Neural machine translation by jointly learning to align and translate. *arXiv preprint arXiv:1409.0473*.

2. Bavota, G., Oliveto, R., Gethers, M., Poshyvanyk, D., & De Lucia, A. (2013). Methodbook: Recommending move method refactorings via relational topic models. *IEEE Transactions on Software Engineering*, 40(7), 671-694.

3. Canfora, G., & Cerulo, L. (2005). Impact analysis by mining software repositories. *Journal of Software Maintenance and Evolution: Research and Practice*, 17(4), 217-232.

4. Gethers, M., Dit, B., Kagdi, H., & Poshyvanyk, D. (2013). Integrated impact analysis for managing software changes. In *Proceedings of the 34th International Conference on Software Engineering (ICSE)*, 430-440.

5. Kagdi, H., Gethers, M., & Poshyvanyk, D. (2013). Integrating conceptual and logical couplings for change impact analysis in software. *Empirical Software Engineering*, 18(5), 933-969.

6. Law, J., & Rothermel, G. (2003). Whole program path-based dynamic impact analysis. In *Proceedings of the 25th International Conference on Software Engineering (ICSE)*, 432-442.

7. Le, T. D. B., Lo, D., Le Goues, C., & Grunske, L. (2016). A learning-to-rank based fault localization approach using likely invariants. In *Proceedings of the 25th International Symposium on Software Testing and Analysis (ISSTA)*, 177-188.

8. McIntosh, S., Kamei, Y., Adams, B., & Hassan, A. E. (2014). The impact of code review coverage and code review participation on software quality: A case study of the qt, vtk, and itk projects. In *Proceedings of the 11th Working Conference on Mining Software Repositories (MSR)*, 192-201.

9. Rolfsnes, T., Di Alesio, S., Behnamghader, P., & Arisholm, E. (2016). Generalizing the analysis of evolutionary coupling for software change impact analysis. In *Proceedings of the 13th International Conference on Mining Software Repositories (MSR)*, 201-212.

10. Vaswani, A., Shazeer, N., Parmar, N., Uszkoreit, J., Jones, L., Gomez, A. N., Kaiser, Ł., & Polosukhin, I. (2017). Attention is all you need. *Advances in Neural Information Processing Systems*, 30, 5998-6008.

11. Wen, F., Tzerpos, V., & Holt, R. C. (2004). Comparing software clustering algorithms based on inter-cluster and intra-cluster coupling. In *Proceedings of the 11th Working Conference on Reverse Engineering (WCRE)*, 96-105.

12. Zimmermann, T., Zeller, A., Weissgerber, P., & Diehl, S. (2005). Mining version histories to guide software changes. *IEEE Transactions on Software Engineering*, 31(6), 429-445.

13. [Authors]. (2024). Scalable and Explainable Enterprise Knowledge Discovery Using Graph-Centric Hybrid Retrieval. *arXiv preprint arXiv:2510.10942*.

14. [Authors]. (2024). Keeping Code-Aware LLMs Fresh: Full Refresh, In-Context Deltas, and Incremental Fine-Tuning. *arXiv preprint arXiv:2511.14022*.